# How to maintain compliance among host country employees who are less anxious after strict government regulations are lifted: An attempt to apply conservation of resources theory to the workplace amid the still-unending COVID-19 pandemic

Keisuke Kokubun[1*], Yoshiaki Ino[2], and Kazuyoshi Ishimura[2]

1 Faculty of Brain Healthcare Business Ecosystem, Graduate School of Management, Kyoto University
2 IEWRI Japan Co., Ltd.

* kokubun.keisuke.6x@kyoto-u.jp

**Abstract**

**Design/methodology/approach**

We compared the awareness of 813 people in Wuhan city from January to March 2023 (Wuhan 2023) and 2,973 people in East and South China from February to May 2020 (China 2020) using responses to questionnaires conducted at Japanese local subsidiaries during each period.

**Purpose**

As the coronavirus pandemic becomes less terrifying than before, there is a trend in countries around the world to abolish strict behavioral restrictions imposed by governments. How should overseas subsidiaries change the way they manage human resources in response to these system changes? To find an answer to this question, this paper examines what changes occurred in the mindset of employees working at local subsidiaries after the government's strict behavioral restrictions were introduced and lifted during the COVID-19 pandemic.

**Findings**

The results showed that the analytical model based on conservation of resources (COR) theory can be applied to both China 2020 and Wuhan 2023. However, the relationship between anxiety, fatigue, compliance, turnover intention, and psychological and social resources of employees working at local subsidiaries changed after the initiation and removal of government behavioral restrictions during the pandemic, indicating that managers need to adjust their human resource management practices in response to these changes.

**Originality/value**

This is the first study that compares data after the start of government regulations and data after the regulations were lifted. Therefore, this research proposes a new analytical framework that companies, especially foreign-affiliated companies that lack local information, can refer to respond appropriately to disasters, which expand damage while changing its nature and influence while anticipating changes in employee awareness.

**Keywords**: COVID-19, psychological resources, social resources, anxiety, fatigue, compliance, turnover intention, change, nationality

**Introduction**

COVID-19, which has raged around the world, continues to increase the number of people infected even today while undergoing repeated mutations and weakening. However, for people who are generally healthy and have no underlying health conditions, such as those who work at manufacturing sites, infection with COVID-19 is no longer as terrifying as the threat of death. Some estimates show that the fatality rate of infected people decreased by 96.8% during the 2.5 years from the onset of the pandemic until mid-2022 (Horita & Fukumoto, 2023). But at the same time, COVID-19 continues to be a thorn in the side of managers. This is because employees are forced to be absent from work due to infection, and if such employees are responsible for irreplaceable duties, work progress will be disrupted. Additionally, if multiple people become infected, temporary closures of individual workplaces may be forced. In workplaces where the risk of infection is high, employees cannot work with peace of mind, and blindly participating in infection control measures may exhaust their physical strength or increase their intention to quit their jobs (Kokubun et al., 2022). Therefore, in today's world where there are no uniformly strict regulations by the government, companies need to think about the measures they should take on their own.

At the beginning of the pandemic, everyone felt anxious in the face of an unprecedented and uncertain situation. However, among those who felt similarly anxious, some were willing to cooperate with infection control measures to make their workplaces safer, while others abandoned workplaces where there was a risk of infection and moved to other workplaces. Why does such a difference occur? A useful reference for considering this issue is the conservation of resources (COR) theory proposed by Hobfoll (1989). COR theory is a stress theory that explains the motivations that drive people to maintain current resources and pursue new resources (Hobfoll, 1989). According to COR theory, when employees experience stress or anxiety and do not have the resources to maintain appropriate behavior at work, they will choose behaviors that are undesirable for the organization, such as leaving the workplace to prevent poor mental health (Jung et al., 2021; Lee & Jang, 2020; Modaresnezhad et al., 2021; Probst et al., 2020; Sinclair et al., 2020). However, if companies can keep their workplaces safe by implementing infection control measures, workers will not have to quit their jobs for fear of resource depletion (Falco et al., 2021; Hu et al., 2021). For this reason, previous studies have mainly adopted cross-sectional analysis methods to analyze the relationship between anxiety and outcomes during the COVID-19 pandemic relying on the COR theory. For example, the pain and anxiety caused by environmental changes during the pandemic decreased work performance and increased intentions to quit (Chen et al., 2023; Kakar et al., 2023; Kumar et al., 2021; Obuobisa-Darko & Sokro, 2023; Singh & Kaurav, 2022), and that alternatively available psychological and social resources moderated these relationships (Pradhan et al., 2023; Farroukh et al., 2023; Kokubun et al., 2022).

So, in what form does this trend exist today after the government restrictions on people's behavior have been lifted and infection control measures have been left to individual companies to make their own decisions? If employees no longer feel as anxious about infection as they once did, companies may be forced to take difficult steps to get those employees to cooperate with the corporate infection control measures. Furthermore, for companies such as foreign subsidiaries that are at a disadvantage in obtaining information about the countries in which they operate, these changes are thought to have a more serious impact on corporate management. With this awareness of the issue, this paper applies an analytical model that reflects the attitudes of employees after the start of strict government restrictions at the beginning of the pandemic to employees after these restrictions ended. Through this, we will explore how human resources should be managed as the stages of the regulation change comparing the data of 813 Chinese employees working for a Japanese manufacturing company in Wuhan from January 5 to March 8, 2023 (hereafter, Wuhan 2023) with the data of 2,973 Chinese employees working for Japanese companies in East and South China from February 15 to May 31, 2020 (hereafter, China 2020). Wuhan was chosen as the study site because it is a city that has been severely affected by multiple pandemics, from the time the virus was first discovered in December 2019 until the time restrictions were lifted in January 2023.

2. **Literature review**

COVID-19 and COR theory

While COVID-19 has great destructive power as a disaster, it does not inflict the same damage on all people, but is extremely uneven, making the situation of socially vulnerable people even worse (Debus et al., 2021; Obinna, 2021; Sy et al., 2021). Therefore, since the reality of this disaster became known, many researchers have applied the COR theory model to understand workplace issues during the pandemic. COR theory asserts that psychological and social resources moderate the relationship between disaster-induced stress and stress-induced outcomes (Hobfoll, 2002). Psychological resources are endogenous resources that include factors such as self-efficacy and resilience, which enable employees to cope with adversity and difficulties (Kim et al., 2017). On the other hand, social resources are interpersonal resources that include elements such as trust, norms, and networks, which can activate cooperative behavior and improve social efficiency (Putnam et al., 1994). Previous research has shown, for example, that the stress and job insecurity caused by COVID-19 hurt employee performance and organizational citizenship behavior, and that trust in management and psychological capital moderate the relationship between them (Farroukh et al., 2023; Pradhan et al., 2023). Other studies have shown that perceived organizational support moderates the relationship between COVID-19 stress and burnout (Tham et al., 2023) and that self-efficacy and social support moderate declines in remote worker happiness and engagement (Straus et al., 2023). In this way, psychological and social-related resources have been shown to have the effect of making the worst-case disaster scenario less likely to occur. This is consistent with the COR theory, which argues that disasters have a more severe impact on people with fewer resources than on those with more resources, and is also consistent with the heterogeneity of the impact of COVID-19 on people.

Kokubun et al. (2022) is one of the earliest studies on COVID-19. In this study, they presented a model that integrates COR theory and arguments about the dual nature of anxiety (Strack et al., 2017; Carver and Scheier, 2011; Norem and Chang, 2002) and conducted an analysis using data from Japanese companies in the East and South regions of China. While anxiety has the positive effect of increasing compliance, that is, understanding and participation in infection control measures, it has the negative effect of increasing fatigue and the intention to quit the job. Therefore, if managers are distracted by the good aspects of the former and only think about getting employees to comply with immediate infection control measures, the bad aspects of the latter will become apparent, and the overall result will be worse. At this time, if there are psychological resources in the workplace, employees will not follow infection control measures driven by anxiety but will participate in them based on their sense of self-efficacy and resilience, keeping fatigue and intentions to quit low. Furthermore, when social resources are available in the workplace, employees are more likely to feel safer at work because they have confidence that they are not alone in following infection control measures and that those around them will also do so, leading to lower turnover intentions (Kokubun et al., 2022).

Many previous studies based on COR theory that dealt with COVID-19 from the perspective of human resource management had three problems. First, they treated anxiety and stress as having only negative effects on outcomes. For example, fear of infection has been shown to negatively impact work, home, and health due to increased emotional suppression and lack of fulfillment of psychological needs (Trougakos et al., 2020). Most recently, Perry and his colleagues (2023) included good stress for behavior change and bad stress that only depletes resources in an analytical model to identify influencing factors of work and family stress during remote work, but this type of research is exceptional. Therefore, Kokubun and his colleagues' (2022) study is unique in showing that it is desirable to suppress anxiety even when considering the positive side of increasing compliance, while also considering the negative side of increasing fatigue and turnover intention.

The other point is related to the time axis. Based on the consensus among researchers that COR theory deals with "changes" in resources and stress, there has been an increasing trend in recent years of studies that incorporate time as an element (Halbesleben et al., 2014; Mäkikangas et al., 2010; Xanthopoulou et al., 2009; Straus et al., 2023). Among these, Straus et al. (2023) found that resources such as self-efficacy and social support can prevent declines in outcomes such as well-being and work engagement, based on an analysis using diary data of remote workers during the pandemic. However, to the authors' knowledge, there is no study based on COR theory that deals with the differences in the relationship between resources, stress, and performance after the end of behavioral restrictions during the pandemic.

The last thing to consider is the nationality of the company. Many studies focused on domestic companies, so there is insufficient understanding of the actual situation at foreign subsidiaries during the pandemic. In general, the scale and social impact of a disaster depend on the country's institutions and regulations (Slovic, 1992). Therefore, to minimize the damage caused by disasters, companies need to have a thorough understanding of the country's circumstances, implement effective strategies, and appropriately manage human resources (Oh & Oetzel, 2011). However, local subsidiaries generally do not have detailed plans or human resources to deal with major disaster risks (Kobrin et al., 1980; Oetzel, 2005; Oh & Oetzel, 2011). Therefore, local subsidiaries are considered to have a weakness, even more than domestic companies, in that they have difficulty obtaining information and responding to disasters in the countries in which they operate. These problems seem more serious in Japanese-affiliated local subsidiaries where authority is not actively delegated to local personnel and therefore expatriates tend to make many important decisions (Kokubun & Yasui, 2021; 2023).

Pandemic, Wuhan, and Japanese subsidiaries' response

To understand the situation after the removal of restrictions, this research targets Chinese employees of Japanese manufacturing companies located in Wuhan from January to March 2023. So, let us look at the evolution of COVID-19 and the response of Japanese subsidiaries in Wuhan. Wuhan is an industrial city in Hubei province, with approximately 200 Japanese companies operating there. As is well known, COVID-19 was first discovered in Wuhan in December 2019. The Chinese government imposed a city lockdown in major cities across China, including Wuhan, for two and a half months from January to April 2020 in response to the spread of the infection. Due to the success of this policy, the country continued to maintain its "zero-Covid policy" to prevent infections, and by mid-2020, China succeeded in containing the spread of the pandemic throughout

China. However, in July 2022, infections were confirmed in the city again, leading to another urban lockdown in Wuhan, following Shanghai in June 2022. This second lockdown was not as effective as expected given the highly contagious variants of the virus. As a result, in November, some Japanese companies took measures such as temporarily shutting down their factories because employees living in areas with movement restrictions were unable to come to work.

Under such circumstances, in December 2022, the Chinese government announced the end of the zero-Covid policy considering cost-effectiveness, and from January 2023 onward, it was left to each company to take measures against infection. However, in December, some media reported that the infection was still spreading and the number of deaths was increasing in Wuhan. As a result, many Japanese companies operating there were forced to continue to take infection control measures such as wearing masks, sanitizing hands, and practicing social distancing in the workplace while remaining vigilant against infection, based on various media and on-site interviews conducted by the authors. The current research, conducted under these circumstances, closely reflects the psychological state of employees after the end of the government's strict restrictions on their movements. Therefore, it can be said to have good conditions for comparison with Kokubun et al. (2022), which was conducted in East and South China after the start of movement restrictions in 2020.

## 3. Hypotheses

Kokubun et al. (2022), who dealt with China 2020, established, and verified the following hypothesis.

*H1. Association between emotions (fatigue and anxiety) and behaviors (compliance and turnover)*

H1a. Anxiety is positively related to turnover intention.

H1b. Anxiety is positively related to fatigue.

H1c. Fatigue is positively related to turnover intention.

H1d. Anxiety is positively related to compliance with COVID-19-related measures.

*H2. Association between behaviors (compliance and turnover)*

H2. Compliance with COVID-19-related measures is negatively related to turnover intention.

*H3. Moderation of resources (social and psychological resources) on the association between anxiety and compliance, anxiety and fatigue, and compliance and turnover)*

H3a. Psychological resources weaken the positive relationship between anxiety and fatigue.

H3b. Psychological resources weaken the positive relationship between anxiety and compliance with COVID-19-related measures.

H3c. Social resources strengthen the negative relationship between compliance with COVID-19-related measures and turnover intention.

A review of the basis for the hypothesis and recent research is as follows. First, about H1. COR theory asserts that individuals use finite resources, such as energy and concentration, and that these resources deplete with use (Hobfoll, 1989). Here, resource depletion leads to chronic symptoms such as emotional exhaustion (Maslach & Leiter, 2008). For example, anxiety has been empirically confirmed to consume energy and lead to resource depletion and emotional exhaustion (Cheng & McCarthy, 2018). When employees feel this fatigue, they may seek a safer workplace and increase their turnover intention to avoid further resource depletion and increased fatigue (Howard & Cordes, 2010; Jung et al., 2021; Lee & Jang, 2020; Modaresnezhad et al., 2021; Probst et al., 2020; Sinclair et al., 2020). Recent research based on COR theory conducted during the pandemic also shows that fatigue mediates the relationship between anxiety and performance, suggesting that anxiety exhausts employees by taking away their emotional resources, depleting them of the energy needed for performance (Jawahar et al., 2022; Yin et al., 2023).

However, anxiety has both good and bad sides. For example, anxiety serves as a signal of how different the desired state is from the actual state. Therefore, anxiety can motivate people to take certain actions by increasing their awareness of the risks to be avoided (Strack et al., 2017; Carver & Scheier, 2011; Norem & Chang, 2002; Schwarz & Bless, 1991). Empirical studies also show that emotional risk perception predicts higher safety compliance and participation (Bozo et al., 2009; Xia et al., 2017). Therefore, it is assumed that the greater the anxiety about COVID-19 among workers, the greater their intention to cooperate with COVID-19 countermeasures. From these discussions, the four hypotheses of H1 are derived.

Next, about H2. This hypothesis is derived by turning the story of H1 on its head. In other words, if the workplace becomes safer and there is no need to worry about energy depletion, employees' intention to leave will decrease. During the pandemic, practicing compliance to prevent the spread of COVID-19 will improve workplace safety. Previous research has shown that safety climate is negatively associated with turnover intention (Smith, 2018; Jung et al., 2021). From the above discussion, H2 was derived.

Finally, about H3. According to COR theory, if required resources are not available, they can be replaced by alternative resources. Hobfoll (2002) states that these alternative resources can be classified into social resources and psychological resources. Psychological resources, including self-efficacy and resilience, enable employees to adapt to adversity, cope with difficulties, and function well in the workplace (Kim et al., 2017; Paek et al., 2015; Song et al., 2020). Therefore, previous studies have shown that psychological resources have a positive effect on safety compliance and participation (Eid et al., 2011; Wang et al., 2018) and a negative effect on emotional fatigue (Moyer et al., 2017; Wang et al., 2012). Furthermore, prior research has shown that psychological resources moderate the relationships between variables. For example, psychological resources were shown to attenuate the negative association between stress and participation in workplace safety measures (Wang et al., 2018). Similarly, another study found that psychological resources weakened the relationship between job anxiety and emotional exhaustion (Shoss et al., 2018) and depression (Aguiar-Quintana et al., 2021). Therefore, even during the pandemic, employees with high psychological resources are thought to be able to weaken the impact of anxiety on their health and behavior.

On the other hand, social resources based on trust, norms, and networks activate people's cooperative behavior and facilitate goal achievement (Putnam et al., 1994; Leana & Van Buren, 1999). Therefore, social resources can be expected to strengthen the relationship between compliance and turnover. As mentioned above, compliance should reduce the risk of infection, increase feelings of security, and thus reduce the willingness to quit work. However, whether compliance leads to actual safety depends on the social resources possessed by people in the workplace (Coleman, 1990; Podolny & Baron, 1997). This is because when social resources are scarce, employees may worry whether their co-workers are following infection control measures themselves (Kokubun & Yamakawa, 2021). In such a psychological state, employees will practice compliance with anxiety, and, contrary to superficial behavior, they may look for opportunities to change jobs to avoid resource depletion. From this, Kokubun et al. (2022) proposed three hypotheses for H3.

In this study, we add H4 and H5, which consist of the following four hypotheses, regarding the differences after the removal of regulations compared to after the start of regulations.

*H4. Changes in the influence of anxiety and fatigue on turnover intention*

*H4a. Anxiety has a smaller effect on the intention to leave the job after the regulations are lifted than after the regulations start.*

*H4b. Fatigue has a greater effect on the intention to leave the job after the regulations are lifted than after the regulations start.*

*H5. Changes in the influence of psychological resources and social resources on compliance*

*H5a. Psychological resources have a greater influence on compliance after the regulations are lifted than after the regulations start.*

*H5b. Social resources have a greater influence on compliance after the regulations are lifted than after the regulations start.*

H4a and H4b are due to the following reasons. First, after the regulation begins, employees who feel insecure, that is, lack of safety resources, will be motivated to leave the workplace to prevent further loss of resources, according to COR theory. However, after deregulation, fear becomes less of an engine of action than before, as government decisions are translated into information to employees that COVID-19 is no longer scary. Currently, the engine of action that replaces anxiety is the fatigue that accumulates due to long-term anxiety and energy depletion. The more fatigued employees are, the higher their intention to quit their jobs is to avoid further energy consumption. This story with a timeline in which anxiety consumes resources and leads to fatigue is consistent with COR theory.

H5a and H5b are due to the following reasons. After the restrictions began, the government's decision was conveyed to employees as information that COVID-19 was scary, and the resulting anxiety prompted employees to take measures to prevent infection. However, after restrictions are lifted, employees will interpret COVID-19 as no longer scary. Therefore, anxiety has less of an impact on encouraging employees to take infection control measures than it did after the restrictions began. However, even after restrictions are lifted, the risk of infection does not disappear. To suppress infections in this situation, companies and employees will need to take voluntary infection control measures even more than when restrictions were in place. Therefore, it is thought that the influence of psychological and social resources on compliance will increase after deregulation.

## 4. Research methodology

### 4.1 Data

The questionnaire consisted of attributes, including age, sex, position, and tenure, as well as 50 question items based on a 5-point Likert response scale from 1 (I do not think this way) to 5 (I do think this way). These items are the same as those used by Kokubun et al.

(2022). Of these, 14 items are social resources such as "The company cares about its employees," 11 items are psychological resources such as "I think I can handle various things well even in a mess," 5 items are compliance such as "I would like to cooperate with the hygiene management of the company to prevent COVID-19 infection," 4 items are anxiety such as ``I'm worried about the COVID-19," 4 items are fatigue such as ``I always feel gloomy because of my work," and 2 items are Turnover intentions such as ``Within a half year, I will quit my current job." Each item was translated into Chinese using the back translation method. For other details on the creation process and items of these variables, please refer to Kokubun et al. (2022).

For each variable, 1 to 5 points were assigned to the individual response items of the 5-point Likert response scale, and the average was calculated for easy comparison. Regarding age, 1 to 4 points were assigned to the options of "under 30", "30-39", "40-49", and "50 or older". Similarly, concerning the length of service, 1 to 4 points were assigned to the options "less than 1 year", "1 year to less than 3 years", "3 years to less than 5 years", and "5 years or more".

This questionnaire was distributed to more than 1,000 Chinese employees at a manufacturing company in Wuhan from January 5 to March 8, 2023. A total of 823 employees responded to it online. However, this analysis uses data from 813 employees who answered all the questions. This paper uses this data (Wuhan 2023) and data collected from 2,973 employees (94.7% of them were manufacturing employees) who were working for 26 companies in the eastern and southern areas from February 15 to May 31, 2020 (China 2020), which was provided by one of the authors of Kokubun et al. (2022). This study was approved by the Ethics Committee of IEWRI Japan Co., Ltd. (approval number 2020–01) and was conducted following the institutes' guidelines and regulations.

All participants provided written informed consent before participation and their anonymity was maintained.

## 5. Analysis and findings

All statistical analyses were performed using IBM SPSS Statistics/AMOS Version 26 (IBM Corp., Armonk, NY, USA). Before proceeding to the main analyses, Harman's single-factor analysis was used to check whether the variance in the data could be largely attributed to a single factor, while the confirmatory factor analysis (CFA) was used to test whether the factors were related to the measures. First, the factor analysis indicated that only 36.3 percent of the variance could be explained by a single factor, which was <50 percent. Thus, it was established that the data did not suffer from common method variance (Chang et al., 2010). Next, for CFA, the model fit was evaluated by examining the chi-square ($\chi^2$), comparative fit index (CFI), standardized root mean square residual (SRMR), and root mean square error of approximation (RMSEA). Values above 0.95 are deemed to indicate a good fit for CFI, and values below 0.05 and 0.08 indicate a good fit for RMSEA and SRMR respectively (Byrne, 1994; Hu and Bentler, 1998). Similar to Kokubun et al (2022), it was shown that the 6-factor model ($\chi^2$ (615) = 1073.550, p<0.001; CFI=0.986; RMSEA=0.030, p<0.001, 90% CI=0.027–0.033; SRMR=0.038) fits better than the 1-factor model that added 6 variables ($\chi^2$ (657) = 2735.487, p<0.001; CFI=0.935; RMSEA=0.062, p<0.001, 90% CI=0.060–0.065; SRMR=0.085).

Table 1 shows the results of the descriptive statistics. Looking at the Wuhan 2023 results shown on the right side of the table, the highest score was 4.39 for psychological resources, followed by 4.38 for compliance and 4.08 for social resources. On the contrary, at 1.87, the turnover intention was the lowest, followed by 2.27 for fatigue and 3.36 for anxiety about COVID-19. It can be said that overall positive awareness is high and negative awareness is low. However, looking at the standard deviation, the former is 0.78 to 0.86, while the latter is around 1.14, indicating that the latter has more variation. Therefore, it should be noted that negative consciousness, especially turnover intention, is not so high on average, but the difference between employees is relatively large. These results are also the same for China 2020 shown on the left side of the table.

Next, we will compare China 2020 and Wuhan 2023 by student's t-test from the same table. Social resources are higher in Wuhan 2023 than in China 2020. On the other hand, Fatigue, Anxiety, and Compliance are lower in Wuhan 2023 than in China 2020. The results were the same in an analysis of covariance (ANCOVA) that controlled for demographic variables age, tenure, manager, and sex (available upon request). These findings suggest that compared to workplaces after the start of restrictions, fatigue and anxiety are reduced in workplaces after restrictions are lifted, and people are less conscious of following workplace infection control measures, while social resources are improved. There are also significant differences between the two groups in Age, Tenure, and Manager (all at the 0.1% level). However, these differences in demographic variables are thought to reflect the uniqueness of this company, as Wuhan 2023, unlike China 2020, targeted a single company. Table 2 shows the results of the correlation analysis conducted for each group. The bottom left of the table shows the results for China 2020, and the top right shows the results for Wuhan 2023.

Table 1 Mean value of each variable and comparison between groups

| | China 2020 | | | Wuhan 2023 | | | |
|---|---|---|---|---|---|---|---|
| | α | Mean | SD | α | Mean | SD | t |
| Social resources | 0.952 | 3.922 | 0.940 | 0.963 | 4.081 | 0.861 | 4.369*** |
| Psychological resources | 0.938 | 4.346 | 0.762 | 0.965 | 4.385 | 0.762 | 1.273 |
| Fatigue | 0.872 | 2.508 | 1.169 | 0.914 | 2.266 | 1.136 | 5.343*** |
| Anxiety | 0.753 | 3.630 | 1.121 | 0.782 | 3.364 | 1.135 | 5.931*** |
| Compliance | 0.914 | 4.639 | 0.707 | 0.874 | 4.380 | 0.781 | 9.023*** |
| Turnover intention | 0.923 | 1.812 | 1.126 | 0.953 | 1.869 | 1.140 | 1.275 |
| Age | - | 1.960 | 0.785 | - | 2.130 | 0.780 | 5.582*** |
| Tenure | - | 2.760 | 1.134 | - | 2.530 | 1.182 | 4.924*** |
| Manager | - | 0.050 | 0.219 | - | 0.020 | 0.130 | 4.133*** |
| Sex | - | 1.620 | 0.485 | - | 1.610 | 0.489 | 0.902 |

Note(s): n = 2973 for China 2020 and n = 813 for Wuhan 2023.

***$p < 0.001$, **$p < 0.01$, *$p < 0.05$. α: the reliability coefficients, t: student's t-test.

Table 2  Results of correlation analysis

| | | 1 | 2 | 3 | 4 | 5 | 6 | 7 | 8 | 9 | 10 |
|---|---|---|---|---|---|---|---|---|---|---|---|
| 1 | Social resources | | 0.639** | -0.408*** | -0.040 | 0.656*** | -0.387*** | 0.160*** | 0.107** | 0.130*** | 0.154*** |
| 2 | Psychological resources | 0.578*** | | -0.237*** | 0.087* | 0.743*** | -0.232*** | 0.153*** | 0.0360 | 0.083* | 0.065 |
| 3 | Fatigue | -0.394*** | -0.203*** | | 0.391*** | -0.220*** | 0.487*** | -0.121** | -0.074* | -0.066 | -0.073* |
| 4 | Anxiety | -0.028 | 0.148*** | 0.353*** | | 0.090* | 0.183** | -0.005 | -0.005 | -0.170*** | 0.090** |
| 5 | Compliance | 0.471*** | 0.674*** | -0.082*** | 0.250*** | | -0.258*** | 0.129*** | 0.079* | 0.093** | 0.062 |
| 6 | Turnover intention | -0.376*** | -0.243*** | 0.424*** | 0.202*** | -0.209*** | | -0.264*** | -0.268*** | -0.101** | -0.170*** |
| 7 | Age | 0.154*** | 0.129*** | -0.138*** | -0.062** | 0.082*** | -0.184*** | | 0.343*** | 0.038 | 0.309*** |
| 8 | Tenure | 0.014 | 0.033 | 0.024 | -0.001 | 0.083*** | -0.113*** | 0.421*** | | 0.149*** | 0.213*** |
| 9 | Manager | 0.074*** | 0.059*** | -0.074*** | -0.117*** | 0.048** | -0.079*** | 0.182*** | 0.167*** | | -0.087* |
| 10 | Sex | 0.047* | 0.019 | -0.013 | 0.079*** | 0.087*** | -0.100*** | 0.099*** | 0.070*** | -0.141*** | |

Note(s): n = 2973 for China 2020 (lower left) and n = 813 for Wuhan 2023 (upper right). ***p < 0.001, **p < 0.01, *p < 0.05.

The figures are the correlation coefficients.

Table 3 shows the results of a simultaneous multi-population analysis conducted to examine the differences in the magnitude of the paths between variables in both groups. First, the negative path from anxiety to turnover intention is significantly larger in China 2020 than in Wuhan 2023 at the 1% level, while the negative path from fatigue to turnover intention is significantly larger in China 2020 than in Wuhan 2023 at the 1% level. These results indicate that anxiety about COVID-19 had a greater impact on intention to leave the job after the government's behavioral restrictions began, and fatigue had a greater effect after the behavioral restrictions were lifted, supporting H4a and H4b. On the other hand, the path to compliance for psychological resources and social resources was shown to be smaller in China 2020 than in Wuhan 2023 at the 0.1% level. These results indicate that participation in infection control measures was more influenced by resources after restrictions were lifted than after restrictions began, supporting H5a and H5b.

Related to F5a and H5b, the negative path of interaction variable between anxiety and psychological resources to compliance is significantly larger in China 2020 than in Wuhan 2023 at the 5% level. This indicates that the effect of psychological resources on suppressing anxiety that drives people to take infection control measures was greater after the restrictions began. This suggests that while resources, especially psychological resources, had a more indirect role in suppressing the impact of anxiety on compliance after the onset of regulation, they had a more direct effect after deregulation encouraging compliance practices.

Let us also look at the relationship between demographic variables and main variables. In the path from sex to compliance, China 2020 is significantly larger than Wuhan 2023 at the 1% level. This shows that women were more proactive than men in cooperating with COVID-19 infection control measures. Similarly, the path from tenure to fatigue is significantly larger for China 2020 than for Wuhan 2023 at the 5% level. Additionally, although there are no significant group differences, the path from tenure to compliance is significant at the 0.1% level in China 2020, but not significant at the 5% level in Wuhan 2023. These results indicate that during the pandemic, long-term employees were more proactive in taking infection control measures than new employees, and at the same time, they were also experiencing greater fatigue.

Table 3 Between-group comparison of paths

| Path | | | Estimate | | t |
|---|---|---|---|---|---|
| | | | China 2020 | Wuhan 2023 | |
| Anxiety | ---> | Compliance | 0.162*** | 0.127*** | 0.794 |
| Sex | ---> | Compliance | 0.045*** | -0.035 | 3.086** |
| Tenure | ---> | Compliance | 0.050*** | 0.026 | 0.918 |
| Social resources | ---> | Compliance | 0.129*** | 0.297*** | 5.728*** |
| Psychological resources | ---> | Compliance | 0.308*** | 0.379*** | 2.597*** |
| Anxiety × Psychological resources | ---> | Compliance | -0.673*** | -0.444*** | 2.107* |
| Anxiety | ---> | Fatigue | 0.346*** | 0.405*** | 1.260 |
| Social resources | ---> | Fatigue | -0.351*** | -0.369*** | 0.938 |
| Psychological resources | ---> | Fatigue | -0.090*** | -0.078* | 0.332 |
| Anxiety × Psychological resources | ---> | Fatigue | -0.119*** | -0.139*** | 0.837 |
| Age | ---> | Fatigue | -0.077*** | -0.046 | 0.904 |
| Tenure | ---> | Fatigue | 0.059*** | -0.023 | 2.415* |
| Compliance | ---> | Turnover intention | -0.275*** | -0.190*** | 1.659 |
| Sex | ---> | Turnover intention | -0.079*** | -0.054 | 0.746 |
| Fatigue | ---> | Turnover intention | 0.241*** | 0.327*** | 2.306* |
| Anxiety | ---> | Turnover intention | 0.114*** | 0.008 | 2.817** |
| Tenure | ---> | Turnover intention | -0.073*** | -0.177*** | 2.842** |
| Age | ---> | Turnover intention | -0.051** | -0.105*** | 1.494 |
| Social resources | ---> | Turnover intention | -0.219*** | -0.197*** | 0.033 |
| Social resources × Compliance | ---> | Turnover intention | -0.260*** | -0.270*** | 0.305 |

Note(s): n = 2973 for China 2020 and n = 813 for Wuhan 2023. $^{***}p < 0.001$, $^{**}p < 0.01$, $^{*}p < 0.05$. t: Student's t value.

The figures are standardized coefficients.

In the above analysis, the model for China 2020 was used, so the goodness of fit for Wuhan 2023 is low. Therefore, we conducted a new path analysis using only Wuhan 2023 data by deleting paths that were not significant and created a model that better reflects the psychological state of the workplace after the abolition of regulations. Figure 1 and Table 4 show the results of path analysis. In the analysis, modification indices were used to improve the model fit. Among demographic variables, only the path from age to turnover intention became significant (omitted in the figure). Regarding the relationship between the main variables, the only difference from China 2020 shown by Kokubun et al. (2022) is that the path from anxiety to turnover intention has disappeared. Therefore, in China 2020, all eight hypotheses from H1 to H3 were supported, and in Wuhan 2023, seven hypotheses except H1a were supported. This indicates the high applicability of Kokubun et al (2022)'s model in different pandemic stages.

As shown in Kokubun et al. (2022), here too, the relationship between anxiety and turnover intention is complicated. Anxiety has the effect of lowering turnover intention through an increase in COVID-19 compliance ($\beta=0.13 \times -0.19 = -0.02$). However, at the same time, it has the effect of indirectly increasing turnover intention through the fatigue increase ($\beta=0.40 \times 0.33 = 0.13$). As a result, the overall effect of anxiety on willingness to leave was positive ($\beta=0.11$), indicating that anxiety enhanced the willingness to leave the job. However, in Wuhan 2023, the overall effect of anxiety on willingness to leave was a little smaller than in China 2020, because there is no longer a direct path from anxiety to intention to leave (in China 2020, the overall effect of anxiety on willingness to leave was β=0.13). However, it can be said that a common trend after the introduction of the regulation and after its abolition is that methods that arouse anxiety end up increasing the intention to quit the job.

Figure 1 Results of path analysis

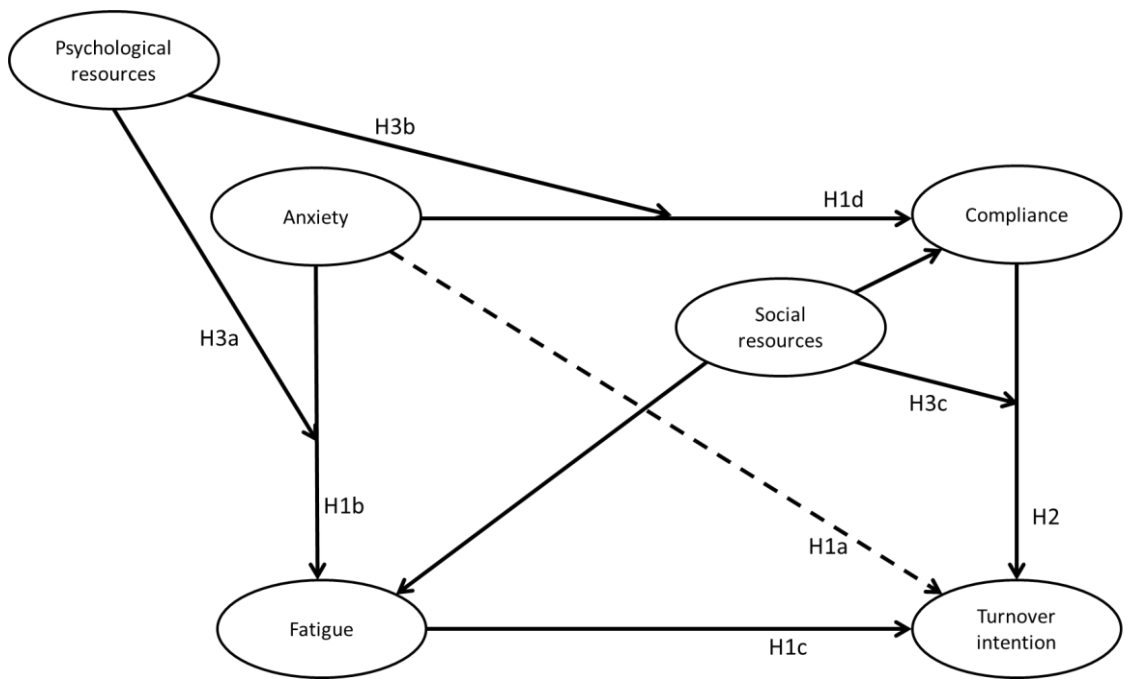

Linear paths are significant at the 0.1 % level. Dashed line paths were not significant even at the 5% level and were therefore excluded from the final model. Goodness-of-fit indices: $\chi^2$ = 40.57, df = 28, root mean square error of approximation (RMSEA) = 0.024, probability of close fit (PCLOSE) = 0.999, goodness of fit index (GFI) = 0.992, adjusted goodness of fit index (AGFI) = 0.977, normed fit index (NFI) = 0.987, comparative fit index (CFI) = 0.996. n = 813.

Table 3   Results of path analysis

| Path | | | Estimate |
|---|---|---|---|
| Anxiety | ---> | Compliance | 0.125 |
| Social resources | ---> | Compliance | 0.293 |
| Psychological resources | ---> | Compliance | 0.379 |
| Anxiety × Psychological resources | ---> | Compliance | -0.446 |
| Anxiety | ---> | Fatigue | 0.404 |
| Social resources | ---> | Fatigue | -0.377 |
| Psychological resources | ---> | Fatigue | -0.083 |
| Anxiety × Psychological resources | ---> | Fatigue | -0.141 |
| Compliance | ---> | Turnover intention | -0.185 |
| Fatigue | ---> | Turnover intention | 0.330 |
| Tenure | ---> | Turnover intention | -0.181 |
| Age | ---> | Turnover intention | -0.116 |
| Social resources | ---> | Turnover intention | -0.203 |
| Social resources × Compliance | ---> | Turnover intention | -0.272 |

Note(s): The numbers in the table are standardized path coefficients. All paths are significant at the 0.1 % level.

Correlation between variables is omitted (available upon request).

To further understand the significance of the interaction terms, in Figure 2 and Figure 3, the data were divided into a group with high psychological resources and a group with low psychological resources. The horizontal axis shows the group with high anxiety and the group with low anxiety, and the vertical axis shows compliance in Figure 2 and fatigue in Figure 3. In Figure 4, the data is divided into a group with high social resources and a group with low social resources; the horizontal axis shows the group with high compliance and the group with low compliance, and the vertical axis shows turnover intention. The criterion for high and low is whether the score is 1 SD higher or lower than the average, following the recommendation of Aiken et al (1991).

Figure 2 The moderating effect of psychological resources between anxiety and compliance

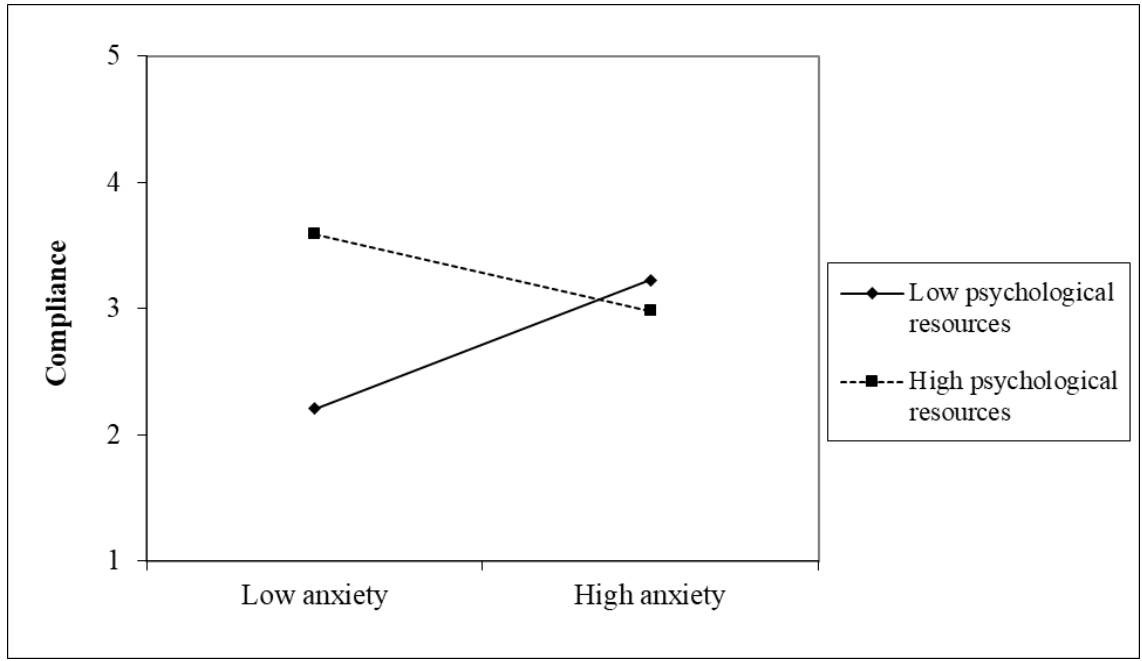

Figure 3 The moderating effect of psychological resources between anxiety and fatigue

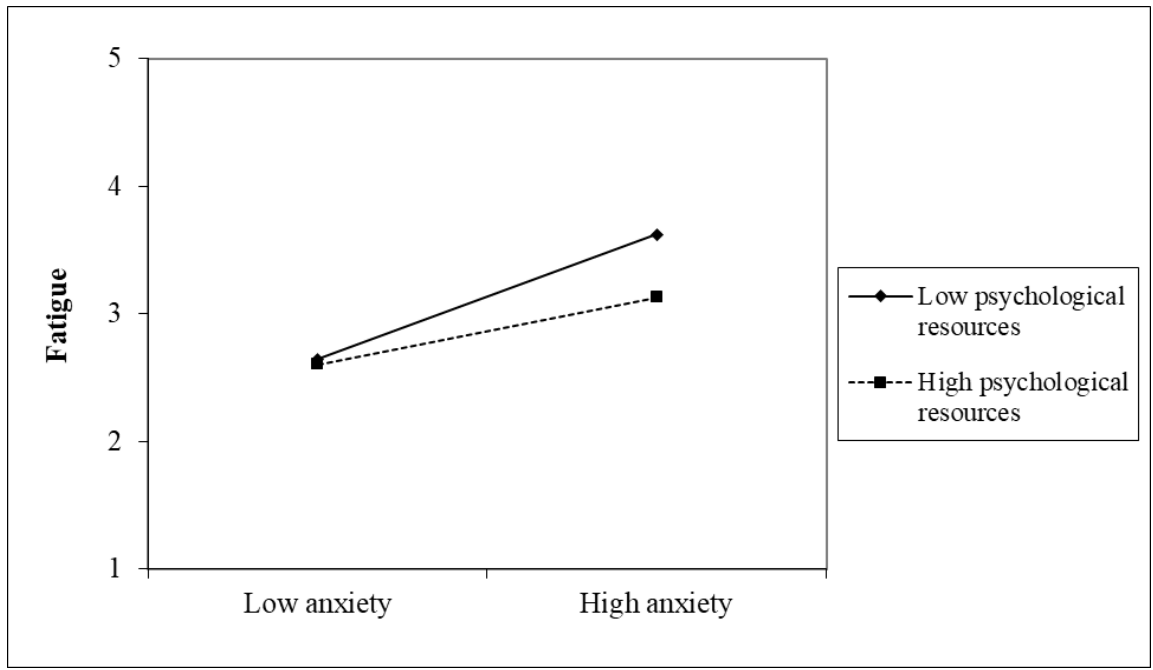

Figure 4 The moderating effect of social resources between compliance and turnover intention

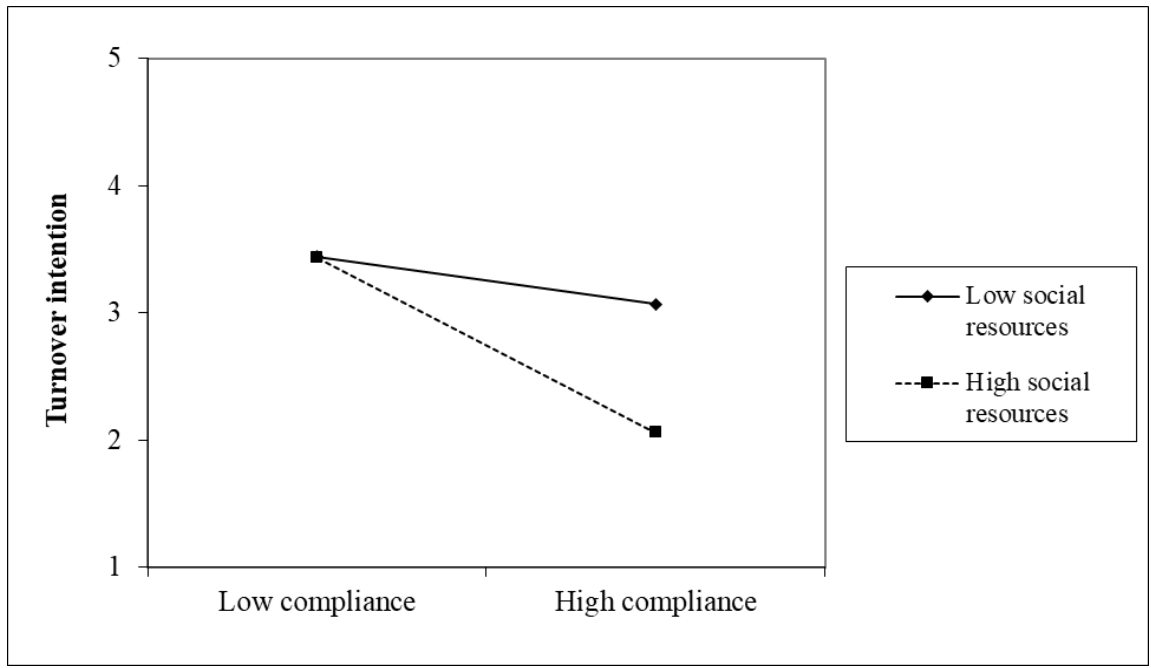

**Discussion**

      In this paper, we first demonstrated that the three points confirmed in data after the start of behavioral restrictions under the COVID-19 pandemic (China 2020), which was found in Kokubun et al. (2022), can also be applied without many changes to the data after restrictions are lifted (Wuhan 2023). First, the more anxious employees are, the more tired they are and the more likely they are to quit their jobs. At the same time, more anxious employees are more likely to be compliant. These are the good and bad ways of inciting fear into submission. Second, compliance lowers turnover intentions. This is because infection control practices improve workplace safety, which reduces the benefits of moving to another workplace due to concerns about reduced resources. Third, psychological resources weaken the relationship between anxiety and fatigue or anxiety

and compliance. Additionally, social resources weaken the relationship between compliance and intention to quit. The former means that people with abundant psychological resources are less likely to feel fatigued even when they are anxious and can practice compliance effectively without arousing anxiety. On the other hand, the latter means that the more social resources an employee has, the more compliance practices will contribute to reducing their intention to quit.

Furthermore, this paper used Kokubun et al.'s (2022) analytical model to clarify changes in employee awareness between 2020, after the government began regulating the industry, and 2023, after the regulations were lifted. First, there are differences in the factors that encourage compliance practices between the start of the regulation and the time after its abolition. In other words, it was shown that social and psychological resources had a greater impact on compliance after the regulation was abolished than after it was implemented. On the other hand, it was also shown that the moderation effect of psychological resources on the path from anxiety to compliance was smaller after the regulation was abolished than after its initiation. After deregulation, compared to after the start of regulations, the role of resources, especially psychological resources, changed from an indirect role of mitigating the impact of anxiety on compliance to directly increasing compliance.

Once restrictions begin, government decisions will be communicated to employees as information that the pandemic is scary. At this stage, anxiety becomes the main driver for compliance practices to prevent the decline of workplace safety resources. However, anxiety can also hinder correct information processing (Cheng and McCarthy, 2018; Easterbrook, 1959). Therefore, infection control measures that rely solely on anxiety are not very effective. On the other hand, after restrictions are lifted, the government's decision will convey to employees the information that COVID-19 is less scary than it was before. As a result, the psychological resources available to deter blind compliance become less effective. However, if compliance decreases due to decreased anxiety, infection becomes more likely to occur. To prevent this problem, regardless of the level of anxiety, compliance will be practiced by making use of resources even after regulations are abolished. The decrease in the moderating effect of psychological resources on the relationship between anxiety and compliance and the increase in the magnitude of the direct path from psychological and social resources to compliance during the period from the start of regulation to the end of regulation is due to a reasonable change for conservation of resources.

Because COR theory is a theory that focuses on stress, it is interesting that fatigue and anxiety showed opposite results in changes from the time the regulation was introduced to the time it was abolished. In other words, the effect of fatigue on turnover intention is greater after the regulation is abolished than after the regulation starts, and conversely, the effect of anxiety on turnover intention is smaller after the regulation is abolished than after the regulation commences. This indicates that there is a temporal difference in the effects of the two on turnover intention and that actions to calm anxiety led to increased resource exhaustion and fatigue over time. This is consistent with the proposed COR theory.

Although it is not directly related to the hypothesis presented in this paper, let us give an overview of changes in the average value. First, compared to China 2020, Wuhan 2023 showed lower levels of anxiety, fatigue, and compliance. This indicates that employees were more anxious and fatigued after the restrictions began than after the restrictions were lifted and that they were therefore more proactive in cooperating with workplace infection control measures. This is consistent with previous research (Chen et al., 2022) that showed that anxiety, depression, and stress among the general population tended to decline over time in Wuhan during the city lockdown. This is consistent with previous research that showed that awareness of compliance with government infection control measures tended to decrease from Wave 1 to Wave 2 (Galasso et al., 2020). After the restrictions began, it is thought that employees were worried about COVID-19 as it was conveyed to them as something mysterious and scary enough to warrant regulation. Therefore, it is thought that compliance awareness and fatigue were high. On the other hand, after restrictions are lifted, employees will feel less anxious than they were at the beginning of the pandemic, as information about the weakening of the virus and government decisions will convey to employees that COVID-19 is no longer scary. Therefore, it is thought that fatigue was low and it was difficult to maintain a high level of compliance awareness.

In addition, social resources and psychological resources showed contrasting results in terms of changes in mean values. In other words, in terms of social resources, Wuhan 2023 is statistically significantly larger than China 2020, but in terms of psychological resources, there is no difference between the two. These findings suggest that after the regulation began, social resources decreased while psychological resources were maintained. Among these, the decline in social resources is related to the fact that it has become difficult to maintain relationships of trust with organizations amidst the chaos of the pandemic and that face-to-face interactions with superiors and colleagues have decreased due to the implementation of remote work. Previous research has shown that it is more difficult to implement social distancing in manufacturing sites, where work requires teamwork, standing work, and work in harsh environments such as high temperatures (Kokubun, 2022). For example, this may have led to less interaction between white-collar employees, who are more likely to work remotely, and blue-collar employees, who are less able to do so, leading to a decline in social resources.

Finally, let us consider the path of demographic variables. This study showed that women were more proactive than men in cooperating with infection control measures after the restrictions began. This may be related to the fact that women are generally more concerned about their health than men (Ek, 2015). Using longitudinal data from OECD countries during the pandemic, Galasso et al. found that women were more likely than men to perceive COVID-19 as a threat to their health and to comply with government infection control measures (Galasso et al., 2020). Women's higher sense of crisis may be related to the fact that the decline in psychological well-being during the pandemic has generally been greater for women than for men, as previous studies have consistently shown (Meyer et al., 2021). These previous studies suggest that women, who are generally more susceptible to health problems due to pandemics, were more proactive in complying than men to protect workplace resources, which is consistent with the results of the current study.

On the other hand, the fact that employees who had worked for longer periods after the start of the regulations were more fatigued compared to after the regulations were abolished can be interpreted as follows. Long-term employees, in other words, are employees who have been accustomed to traditional ways of working for a long time. Therefore, in the face of unprecedented disasters and changes in work styles, it is thought that a great deal of physical strength was consumed in adapting. They also likely had relatively greater responsibility in the workplace and were in a position to mentor less experienced employees. This is also thought to have had the effect of draining their physical strength.

Thus, after the introduction of the restrictions, compared to after the removal of the restrictions, women were more proactive in taking infection control measures, and those who had worked for a longer period were more fatigued. These are helpful tips to overcome disasters like the COVID-19 pandemic. For example, in workplaces where there is no communication between men and women due to differences in their roles, or where women have little say, there is a possibility that awareness of compliance will be weakened and infection control measures will not be implemented appropriately. Additionally, unless attention is paid to fatigue among long-term employees and appropriate measures are taken, veteran employees who are loyal and highly skilled will be more likely to leave, harming the company. The results of this analysis show that the daily activities of listening to the voices of diverse human resources and incorporating them into management become even more important in times of disaster.

**Implication**

In this study, we conducted a simultaneous multi-population analysis by inputting data on employee attitudes after the abolition of regulations into an analytical model developed based on data on employee attitudes after the start of regulations during the COVID-19 pandemic. Based on the COR theory, we then clarified the magnitude of the path from anxiety, fatigue, and compliance to intention to quit, and the changes in moderation by psychological and social resources. Several previous studies have dealt with changes in employee attitudes during the pandemic, but as far as the authors know, this is the first study that compares data after the start of government regulations and data after the regulations were lifted. Therefore, this research proposes a new analytical framework that companies, especially foreign-affiliated companies that lack local information, can refer to respond appropriately to disasters, which expand damage while changing its nature and influence while anticipating changes in employee awareness.

Kokubun et al. (2022) presented a model that combines the COR theory with the dual nature of anxiety and conducted an analysis using employee awareness data after the start of regulations. As a result, it has been argued that infection control measures that are implemented by inciting anxiety can also increase fatigue and intentions to quit, and reduce organizational strength. On the other hand, this study shows that Kokubun et al.'s (2022) model can be applied to post-deregulation employee attitude data without major changes. In addition, the factor influencing turnover intention has changed from anxiety to fatigue, and at the same time, the role of psychological and social resources has changed from an indirect role of moderating the relationship between anxiety and compliance to a direct role of increasing compliance. These results demonstrate that the analytical model presented by Kokubun et al. (2022) is robust under different pandemic stages, and at the same time, different strategies are required depending on the stage. During the pandemic, companies were able to get employees to participate in infection control measures simply by inciting anxiety, but once restrictions are lifted, in addition to anxiety, the lack of psychological and social resources will motivate employees to take infection control measures. However, unlike anxiety, these resources cannot be developed overnight. Foreign-affiliated companies with large differences between national and organizational cultures will have had a harder time implementing infection control measures after restrictions were lifted. I hope that this research will provide good implications for the future activities of overseas subsidiaries of various nationalities.

Study limitations and suggestions for future research.

This study has some limitations. First, the biggest problem is that we compared China in 2020 and Wuhan in 2023 after the restrictions began and after the restrictions were lifted, respectively. The former data was collected from 26 companies operating in East and South China regions, while the latter data was collected from one company. Therefore, there may be regional influences on the results. In the future, it is recommended to conduct research that includes the same number of regions and companies for comparison and to verify the results of this study. Second, there is the issue of generalizability. This study targeted employees of Japanese companies in China. Therefore, in the future, it will be necessary to verify whether the same applies to employees of companies of other nationalities in other countries. Third, in this study, we used the scale developed by Kokubun et al. (2022) instead of a general psychological scale. Therefore, it is necessary to verify reproducibility using a more general scale.

**Conclusion**

What kind of changes has the coronavirus pandemic brought about in the mindset of local employees working at foreign companies? This research was conducted to understand this situation and use it to help local subsidiaries take measures against infection. The same survey that was conducted targeting Japanese companies in East and South China after the strong government restrictions began was conducted among Japanese manufacturing companies in Wuhan after the restrictions were lifted. As a result, the model based on COR theory that was shown using data from China after the start of restrictions (China 2020) can be applied without major changes even when using data from Wuhan after restrictions are lifted (Wuhan 2023). However, there were some significant differences between the two in terms of path size. In other words, the results showed that: (1) anxiety had a smaller effect on the intention to leave the job, and fatigue had a greater effect on the intention to leave the job; also showed that (2) psychological and social resources had a large influence on compliance after the regulation was lifted compared to after the start of the regulation. These results suggest that the kind of anxiety-provoking infection control measures used during the pandemic will be less effective once restrictions are lifted, and efforts to increase workplace resources while focusing on employee health management are needed from a long-term perspective.

**Data Availability**

The data that support the findings of this study are available from the corresponding author, KK, upon reasonable request.

**Author Contributions**

KK performed the data analysis, wrote the main manuscript text, prepared the figures and tables, and conducted the supervisory work. YI and KI were responsible for project administration. All authors reviewed, edited, and agreed to the published version of the manuscript.

**Conflicts of Interest**

YI and KI were employed by IEWRI Japan Co., Ltd. The remaining author declares that the research was conducted in the absence of any commercial or financial relationships that could be construed as a potential conflict of interest. The raw data supporting the conclusions of this manuscript will be made available by the authors, without undue reservation, to any qualified researcher.